\documentclass[sigconf,screen]{acmart}

\acmConference[Draft Manuscript]{Draft Manuscript}{August 2026}{}

\usepackage{amsmath}
\usepackage{microtype}
\usepackage{wrapfig}
\usepackage{booktabs} 
\usepackage{makecell} 

\usepackage{amsfonts}
\usepackage[linesnumbered,commentsnumbered,ruled,vlined]{algorithm2e}
\usepackage{algorithmic}

\SetCommentSty{mycommfont}

\usepackage{mathpartir}

\usepackage{pifont}

\newcommand{\cmark}{\ding{51}} 
\newcommand{\xmark}{\ding{55}} 

\SetKwRepeat{Do}{do}{while}

\usepackage{multicol}
\usepackage{latexsym}
\usepackage{multirow}
\usepackage{graphicx}
\usepackage{threeparttable}
\usepackage{tikz}
\usetikzlibrary{arrows, automata, positioning,shapes, patterns}

\usepackage{circuitikz}

\usepackage{verbatim}

\usepackage{tcolorbox}
\usepackage{centernot}

\newif\ifshowcomments
\showcommentsfalse

\ifshowcomments

\newcommand{\yao}[1]{\mytodoblue{[yao: #1]}}
\newcommand{\shi}[1]{\mytodocyan{[shi: #1]}}
\else
\newcommand{\yao}[1]{}
\newcommand{\shi}[1]{}
\fi

\newcommand{\mytodoblue}[1]{\textcolor{blue}{\ding{46}~{\sf}~#1}}

\newcommand{\mytodocyan}[1]{\textcolor{cyan}{\ding{46}~{\sf}~#1}}

\usepackage[utf8]{inputenc}

\newcommand{\ToolName}{Phoenix}

\usepackage{listings}
\lstdefinelanguage{tiny-lang}
{
	morekeywords={define},
	sensitive=false,
	morecomment=[l]{;}
}
\usepackage{pgfplots}
\pgfplotsset{compat=1.9}
\usepackage{balance}
\usepackage{adjustbox}
\usepackage{proof}
\usepackage{stmaryrd}

\usepackage{url}
\hypersetup{
	colorlinks=true,
	linkcolor=blue,
	citecolor=violet,
	filecolor=magenta,
	urlcolor=blue,
}

\usepackage{cleveref}
\crefname{section}{§}{§§}
\Crefname{section}{§}{§§}

\begin{document}

\title{\ToolName: A Modular and Versatile Framework \\ for C/C++ Pointer Analysis}

\author{Peisen Yao}
\affiliation{%
  \institution{The State Key Laboratory of Blockchain and Data Security, Zhejiang University}
  \city{Hangzhou}
  \country{China}}
\email{pyaoaa@zju.edu.cn}

\author{Zinan Gu}
\affiliation{%
  \institution{The State Key Laboratory of Blockchain and Data Security, Zhejiang University}
  \city{Hangzhou}
  \country{China}}
\email{guzinan1998@zju.edu.cn}

\author{Qingkai Shi}
\affiliation{%
  \institution{The State Key Laboratory for Novel Software Technology, Nanjing University}
  \city{Nanjing}
  \country{China}}
\email{qingkaishi@nju.edu.cn}

\renewcommand{\shortauthors}{Yao, Gu, and Shi}

\begin{abstract}
We present \ToolName, a modular pointer analysis framework for C/C++ that unifies multiple alias analysis algorithms behind a single, stable interface. \ToolName\ addresses the fragmentation of today’s C/C++ pointer analysis ecosystem by cleanly separating IR construction, constraint generation, solver backends, and client-facing queries—making analyses easy to compare, swap, and compose while exposing explicit precision--performance trade-offs. 
We evaluate two comparisons on 28 GNU coreutils programs: \ToolName-FICI versus SVF-FICI, where both analyses are flow-insensitive and context-insensitive, and \ToolName-FSCS versus SVF-FSCI, where the former is flow- and context-sensitive and the latter is flow-sensitive but context-insensitive. \ToolName-FICI is faster on all benchmarks, with a maximum speedup of 2.88$\times$. For the second comparison, \ToolName-FSCS is faster on 13 benchmarks, slower on 14, and tied on one, with a maximum speedup of 2.91$\times$. 
In production, \ToolName\ serves as the analysis substrate for static analysis and fuzzing tools that have uncovered hundreds of new bugs in open-source software and industrial deployments.
\end{abstract}

\maketitle

\section{Introduction}
\label{sec:intro}

Pointer analysis~\cite{hind2001pointer,hind2000pointer} is a fundamental family of static analyses that estimates the possible values of
pointer variables in a program. Such information 
underpins a wide range of applications, including compiler optimizations~\cite{lattner2007making}, program slicing~\cite{li2016program,sridharan2007thin}, bug detection~\cite{sui2012static,shi2018pinpoint,tripp2009taj,YaoFalcon24}, change-impact analysis~\cite{Orso:2004:ECD:998675.999453}, and program verification~\cite{wang2017partitioned,fink:typestate:issta,gurfinkel2015seahorn,kuderski2018teadsa}. 
A more precise pointer analysis is typically favored by clients, as it implies, e.g., fewer false alarms, more accurate code navigation, and greater optimization potential.

For managed languages such as Java, pointer analysis has matured into a rich ecosystem of reusable infrastructure, including WALA~\cite{WALA}, Doop~\cite{DOOP}, Soot~\cite{Soot}, SootUp~\cite{SoopUp}, OPAL~\cite{OPAL}, and Taie~\cite{Taie}.
These analysis frameworks expose stable intermediate representations and shared abstractions, making it practical to (i) implement new analyses without committing to low-level engineering choices, and (ii) compose research prototypes with mature components.

In contrast, effective pointer analysis for C/C++ remains difficult. 
The language exposes a low-level memory model with unrestricted pointer arithmetic, manual memory management, type punning, and complex data-structure idioms, all of which complicate sound and precise reasoning. Despite decades of progress~\cite{smaragdakis2015pointer,hind2001pointer}, the C/C++ tool landscape remains fragmented: frameworks often hard-code a single analysis style, offer limited extensibility, and provide little interoperability across clients and solvers.

\begin{table}[t]
\centering
\caption{Comparison of pointer analyses in static analysis frameworks for C/C++. The ``Eq/Sub'' column indicates whether assignments are modeled with equality or subset constraints; \cmark, \xmark, and ``--'' indicate support, lack of support, and not applicable, respectively.
}
\label{tab:framework-comparison}
\resizebox{\linewidth}{!}
{
\begin{tabular}{l|c|cccc}
\toprule
\textbf{Framework} & \textbf{IR} & \textbf{Eq/Sub} & \textbf{\makecell{Exhaustive\\Context-Sen.}} & \textbf{\makecell{Exhaustive\\Flow-Sen.}} & \textbf{C++} \\
\midrule
SVF        & LLVM & Sub & \xmark & \cmark & \cmark \\
DG         & LLVM  & Sub & \xmark & \cmark & \cmark \\
SeaDSA     & LLVM & Eq  & \cmark & \xmark & \cmark \\
Phasar     & LLVM & --  & --     & --     & \cmark \\
CPAchecker & CFA (from CDT)   & Sub  & \xmark     & \xmark & \xmark \\
\midrule
\textbf{\ToolName} & LLVM IR & Eq, Sub & \cmark & \cmark & \cmark \\
\bottomrule
\end{tabular}
}
\end{table}

Table~\ref{tab:framework-comparison} compares representative static analysis frameworks for C/C++ with respect to their pointer analysis capabilities. LLVM~\cite{lattner2004llvm} includes several built-in alias analyses (e.g., BasicAA), but these are designed primarily for compiler optimizations and offer limited precision. Software model checkers often rely on pointer analysis for memory modeling. For example, SeaHorn~\cite{gurfinkel2015seahorn} and SMACK~\cite{carter2016smack} use unification-based pointer analysis~\cite{lattner2007making,SeaDSA} to derive region-based memory abstractions. CBMC~\cite{kroening2014cbmc} and CPAchecker~\cite{beyer2011cpachecker} implement custom pointer analyses over their respective intermediate representations. These analyses are typically designed for scalability and omit flow- or context-sensitivity.

Dedicated pointer analysis frameworks also exhibit trade-offs. SVF~\cite{sui2016svf} implements Andersen-style pointer analysis with value-flow tracking, but offers limited configurability. For example, it lacks support for (1) exhaustive context-sensitive analysis and (2) exhaustive flow- and context-sensitive analysis.\footnote{SVF has a demand-driven, flow- and context-sensitive pointer analysis.} 
DG~\cite{chalupa2020dg} focuses on dependence representations and includes a flow-sensitive Andersen variant, but does not aim to provide a general, extensible pointer-analysis platform.
Phasar~\cite{schubert2019phasar} targets interprocedural dataflow analyses and typically delegates aliasing to external engines, such as LLVM and SVF.

This fragmentation has practical costs: comparing analyses requires reimplementing them across incompatible infrastructures, and composing or extending tools demands ad hoc integration with internal data structures—hindering reproducibility and portability.

To address these challenges, we follow the \textbf{IDEA} principles, namely \emph{Integrate}, \emph{Diversify}, \emph{Extend}, and \emph{Advance}:
\begin{itemize}
\item \textbf{I}ntegrate: Provide a unified architecture that enables analyses to interoperate via shared abstractions.
\item \textbf{D}iversify: Support multiple analysis paradigms, allowing users to select precision--performance trade-offs per task.
\item \textbf{E}xtend: Offer structured APIs and configuration mechanisms that reduce the effort of developing new analyses.
\item \textbf{A}dvance: Modernize classical pointer analysis techniques via modular implementations.
\end{itemize}

We present \ToolName, a modular pointer analysis framework for C/C++ built within the Lotus program analysis infrastructure~\cite{lotus2025}. \ToolName\ unifies multiple pointer analysis algorithms behind a stable query interface.
 Clients can switch between equality- and subset-based solvers, and between flow- and context-sensitive variants, without modifying client logic.
These capabilities come from three design elements: configurable flow/context variants that expose precision--performance trade-offs, a stable query API that hides solver-specific details, and a modular separation of IR construction, constraint generation, and solving backends.

We use FICI for flow-insensitive, context-insensitive analysis, FSCI for flow-sensitive, context-insensitive analysis, and FSCS for flow-sensitive, context-sensitive analysis. We evaluate two comparisons between \ToolName\ and SVF~\cite{sui2016svf} on 28 GNU coreutils programs. First, \ToolName-FICI is compared with SVF-FICI; both analyses are flow-insensitive and context-insensitive, and \ToolName-FICI is faster on all benchmarks, with a maximum speedup of 2.88$\times$. Second, \ToolName-FSCS is compared with SVF-FSCI, SVF's flow-sensitive but context-insensitive configuration. In this comparison, \ToolName-FSCS is faster on 13 benchmarks, slower on 14, and tied on one, with a maximum speedup of 2.91$\times$.

\ToolName\ has served as a foundation for higher-level analyses, including static bug finding~\cite{sun2024fast,OOPSLA22CFL}, numerical abstract interpretation~\cite{ISSTA25Bit}, and directed fuzzing~\cite{HuangTitan24,HuangBeacon22}.
These applications have uncovered hundreds of previously unknown defects in open-source software and have supported industrial deployments at Huawei. We hope \ToolName\ will serve as a common framework for the program analysis and verification community, fostering reuse, reproducibility, and the rapid development of new techniques.

\section{Pointer Analyses in \ToolName}
\label{sec:method}

\begin{figure}[t]
\centering
\includegraphics[width=\linewidth]
{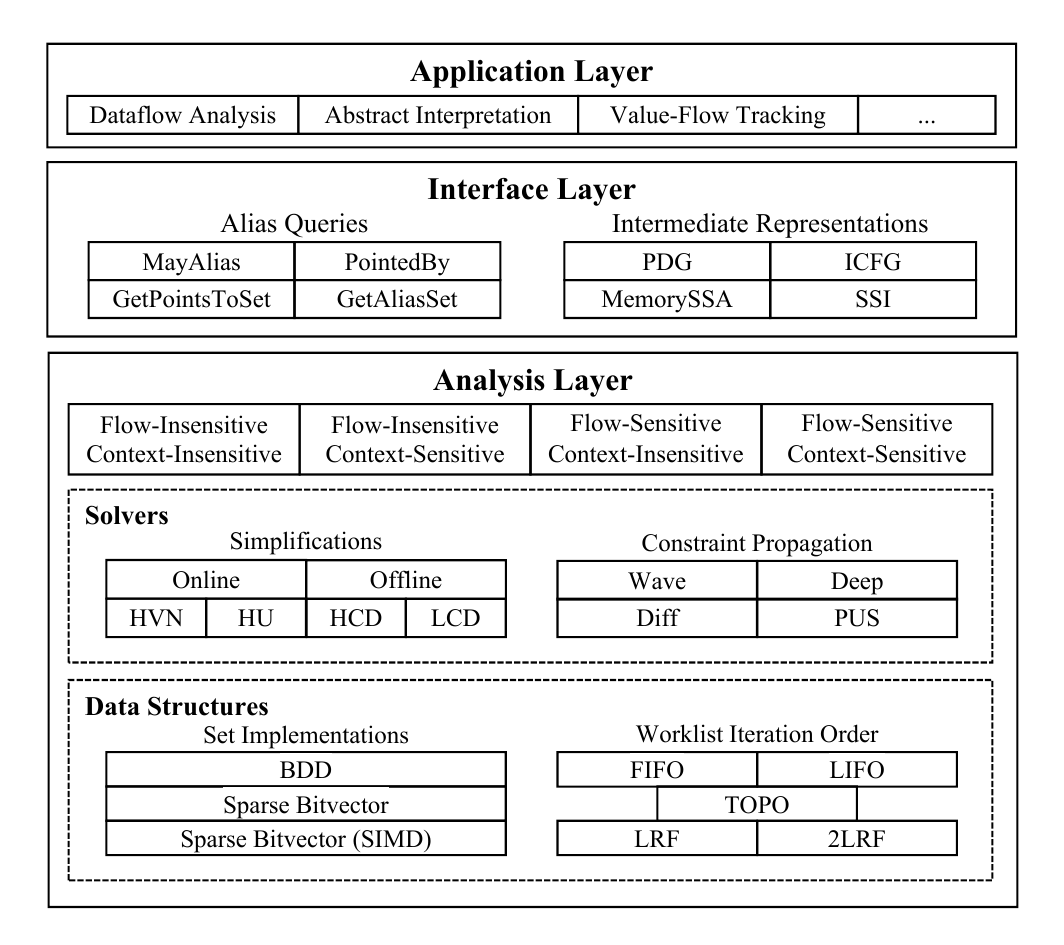}
\caption{\ToolName\ framework architecture.}
\label{fig:architecture}
\end{figure}

As shown in Figure~\ref{fig:architecture},
 \ToolName\ features a modular architecture that unifies diverse pointer analysis algorithms into reusable components: (i) a common constraint
language and IR-to-constraint front-end, (ii) interchangeable solver backends and optimizations, and (iii) standardized result adapters consumed by the query layer.

\begin{table}[t]
\centering
\caption{Configurable techniques and solver choices.
}
\label{tab:techniques}
\resizebox{\linewidth}{!}
{
\begin{tabular}{lll}
\toprule
Technique & Abbrev. & Described in \\
\midrule
\textbf{Offline Simplifications} & & \\

$\square$ Hash-based Value Numbering & HVN & Hardekopf and Lin~\cite{hardekopf2007exploiting} \\
$\square$ HVN with Dereference and Union & HU & Hardekopf and Lin~\cite{hardekopf2007exploiting} \\
\midrule
\textbf{Online Simplifications} & & \\
$\square$ Hybrid Cycle Detection & HCD & Hardekopf and Lin~\cite{hardekopf2007ant} \\
$\square$ Lazy Cycle Detection & LCD & Hardekopf and Lin~\cite{hardekopf2007ant} \\
\midrule
\textbf{Solver} & & \\
$\square$ Wave Propagation & Wave & Pereira et al.~\cite{pereira2009wave}\\
$\square$ Deep Propagation & Deep & Pereira et al.~\cite{pereira2009wave}\\
$\square$ Difference Propagation & Diff & Pearce et al.~\cite{pearce2007efficient} \\
$\square$ Partial Update Solver & PUS & Liu et al.~\cite{liu2022pus} \\
\midrule
\textbf{Worklist Iteration Order} & & \\
$\square$ First In First Out & FIFO & \multirow{4}{*}{Pearce et al.~\cite{pearce2007efficient}} \\
$\square$ Last In First Out & LIFO & \\
$\square$ Least Recently Fired & LRF & \\
$\square$ Topological & TOPO & \\
$\square$ 2-Phase Least Recently Fired & 2LRF & Hardekopf and Lin~\cite{hardekopf2007ant} \\

\bottomrule
\end{tabular}
}
\end{table}

\smallskip 
\noindent   \textbf{Data Structures}. 
Efficient points-to manipulation is critical for scalability.
\ToolName\ decouples the abstract interface of points-to sets from their
concrete representation, and supports multiple set implementations:
\emph{BDD-based} representations for compact storage of large points-to sets, and
\emph{sparse bitvectors} for fast bulk set operations.
In addition, \ToolName\ provides a pluggable backend interface to support
specialized implementations, including a SIMD-accelerated sparse bitvector
backend~\cite{tan2026simd}.

\smallskip 
\noindent  \textbf{Constraint Solving}. 
\ToolName\ provides an interchangeable solver layer with configurable optimizations
summarized in Table~\ref{tab:techniques}.
The key design choice is to treat simplifications and propagation strategies as first-class, composable modules, allowing for the instantiation of a large space of solver configurations without rewriting core logic.

\smallskip 
\noindent   \textbf{Algorithm Portfolio}. \ToolName\ integrates an algorithm portfolio that spans a broad range of
precision--performance trade-offs.
To make these trade-offs explicit and configurable, we categorize algorithms by their semantic abstraction.

\smallskip 
\noindent \emph{Inclusion-Based Analyses.}
Inclusion-based analyses model assignments as directional constraints (e.g., $p \supseteq q$), preserving fine-grained flow of information between pointers.
\ToolName\ includes Andersen analysis as the canonical baseline:
\begin{itemize}
  \item {Andersen Analysis}: A flow-insensitive, context-insensitive, subset-based
  analysis that serves as the baseline in many pointer analysis frameworks.
  \item {Context-Sensitive Variants}: A configurable family of variants
  that refines aliasing across call sites. \ToolName\ supports systematic context abstractions (e.g., 1-CFA, 2-CFA, etc.) to enable controlled studies of precision and cost.
  \item {Flow-Sensitive Variants}: \ToolName\
  can incorporate control-flow awareness into constraint propagation while reusing the same inclusion-constraint semantics.
\end{itemize}


\smallskip
\noindent \emph{Unification-Based Analyses.}
Unification-based analyses trade precision for speed by merging pointers that
are related by assignments, yielding near-linear performance in practice for many
codebases.
\ToolName\ supports unification-style analyses through equality constraints and
implements Dyck-style variants that provide a principled way to recover useful precision for common C/C++ idioms while retaining efficient solving.

\section{The Query Interface of \ToolName}
\label{sec:querying}
\ToolName\ exposes a \emph{unified query interface} that hides algorithm-specific
representations behind a stable set of operations. The key goal is to decouple
client analyses from (i) the choice of pointer-analysis algorithm and (ii) solver- and optimization-level engineering details.
To this end, \ToolName\ adopts an \emph{adapter} design: each analysis backend provides a thin
result adapter that normalizes its output into a common set of query primitives, while the interface layer offers higher-level queries built from these primitives.

\smallskip
\noindent \textbf{Core Pointer-Related Queries}.
Table~\ref{tab:queries} summarizes the common pointer-related queries supported by \ToolName. These operations support a wide range of client analyses—including memory safety verification, information-flow tracking, and concurrency bug detection—and are \emph{composable}: more specialized client logic can be expressed as combinations of these primitives without depending on backend internals.

\begin{table}[t]
\centering
\caption{Pointer analysis queries exposed by the unified abstraction layer.}
\label{tab:queries}
\resizebox{\linewidth}{!}
{
\begin{tabular}{lll}
\toprule
\textbf{Query} & \textbf{Description} & \textbf{Typical Application} \\
\midrule
\texttt{MayAlias($p, q$)} & Whether two pointers may alias & Data race detection \\
\midrule
\texttt{PointedBy($p, o$)} & Whether object $o$ is in the points-to set of pointer $p$ & Escape analysis \\
\midrule
\texttt{GetPointsToSet($p$)} & Memory locations that pointer $p$ may reference & Cast failure analysis \\
\midrule
\texttt{GetAliasSet($v$)} & The set of values that may alias $v$ & Taint analysis\\
\bottomrule
\end{tabular}
}
\vspace{-3mm}
\end{table}

\smallskip
\noindent \textbf{Intermediate Representations}.
Beyond answering direct pointer queries, \ToolName\ provides access to a set of intermediate representations (IRs) that capture control, data, and memory dependencies.
These IRs serve as a shared substrate for downstream analyses:

\begin{itemize}
  \item {PDG}: an interprocedural graph of data and control dependencies,
  supporting clients such as program slicing.
  \item {ICFG}: an interprocedural CFG with call and return edges.
  \item {MemorySSA}: an SSA-style memory representation with implicit use--def
  links for memory-dependence reasoning.
  \item {Static Single Information (SSI)}: extends SSA with $\sigma$ (split) and $\pi$ (merge) nodes
  to support bidirectional information flow, benefiting analyses that mix forward and backward propagation.
\end{itemize}

\section{Applications of \ToolName}
\label{sec:applications}

By exposing a common interface over diverse pointer analysis algorithms, \ToolName\ enables a range of clients that depend on memory reasoning. 
This section outlines both existing use cases (\cref{subsec:uses}) and prospective applications (\cref{subsec:future}).

\subsection{Existing Use Cases}
\label{subsec:uses}
\ToolName\ underpins several advanced analyses developed within the broader Lotus framework, including several prior works by the authors~\cite{sun2024fast,OOPSLA22CFL,ISSTA25Bit,HuangTitan24,HuangBeacon22}:

\begin{itemize}
	\item \emph{Null Pointer Analysis}~\cite{sun2024fast}: \ToolName\ enables scalable detection of null dereference errors by filtering infeasible dereference sites before applying an expensive path-sensitive analysis. The underlying engine also supports other memory safety properties, such as use-after-free and buffer overflows.
	\item \emph{Taint Analysis}~\cite{OOPSLA22CFL}: Interprocedural taint tracking relies on accurate points-to information to propagate tainted data through complex pointer operations. The unified interface allows taint analyses to switch between pointer algorithms based on precision and performance requirements.

	\item \emph{Numerical Analysis}~\cite{ISSTA25Bit}: The cited work introduces an abstract domain combining bit-level precision with word-level reasoning for low-level code. Accurate pointer information is useful for tracking value flows through memory operations.
    \item \emph{Directed Fuzzing}~\cite{HuangTitan24,HuangBeacon22}: We use the pointer information to guide fuzzers in steering input generation toward targeted program regions (e.g., for reproducing crash reports, confirming static analysis results, and regression testing), improving crash discovery for complex C/C++ binaries. 
\end{itemize}

\subsection{Potential Applications}
\label{subsec:future}

\ToolName\ is designed to support both pointer analysis researchers and users. Its modular architecture and standardized interfaces reduce the engineering effort required to build or evaluate pointer-aware analyses. Improvements to underlying pointer algorithms are immediately reflected in all dependent clients, promoting reuse and accelerating research.

\begin{itemize}
    \item \emph{Software Model Checking}:  Several software verifiers, such as SeaHorn~\cite{gurfinkel2015seahorn,kuderski2018teadsa,gurfinkel2017context} and SMACK~\cite{carter2016smack}, rely on alias analysis for modeling memory. \ToolName\ provides a flexible infrastructure for experimenting with alternative aliasing models, enabling verification researchers to explore different precision--performance trade-offs.
    \item \emph{Security Analysis}: An IFDS/IDE~\cite{reps1998program2} interprocedural dataflow analysis framework has been implemented atop \ToolName, enabling security applications such as taint analysis and policy enforcement.
    \item \emph{Semantic Indexing}: Alias information is important for tools that perform program slicing, automated refactoring, and other coding tasks. \ToolName\ enables such tools to reason about memory aliasing and side effects with higher fidelity. 
\end{itemize}

\section{Evaluation}
\label{sec:evaluation}

\noindent \textbf{Analysis Performance}.
We evaluate \ToolName\ against SVF~\cite{sui2016svf}, a state-of-the-art, actively maintained, and widely used pointer analysis framework for C/C++.
Using the notation defined in \cref{sec:intro}, we consider two comparisons. First, \ToolName-FICI is compared with SVF-FICI; both are flow-insensitive and context-insensitive Andersen-style analyses. Second, because SVF does not provide an exhaustive flow- and context-sensitive analysis, \ToolName-FSCS is compared with SVF-FSCI, SVF's flow-sensitive but context-insensitive configuration. \ToolName-FSCS is flow-sensitive and uses  the configuration of 2-CFA for  context sensitivity.

We run all four analysis configurations on a suite of $28$ GNU coreutils programs that exhibit diverse pointer behaviors. 
All experiments are conducted on a Linux server equipped with an Intel Xeon CPU and 512 GB RAM.

\begin{figure}[t]
  \centering
  \includegraphics[width=\linewidth]{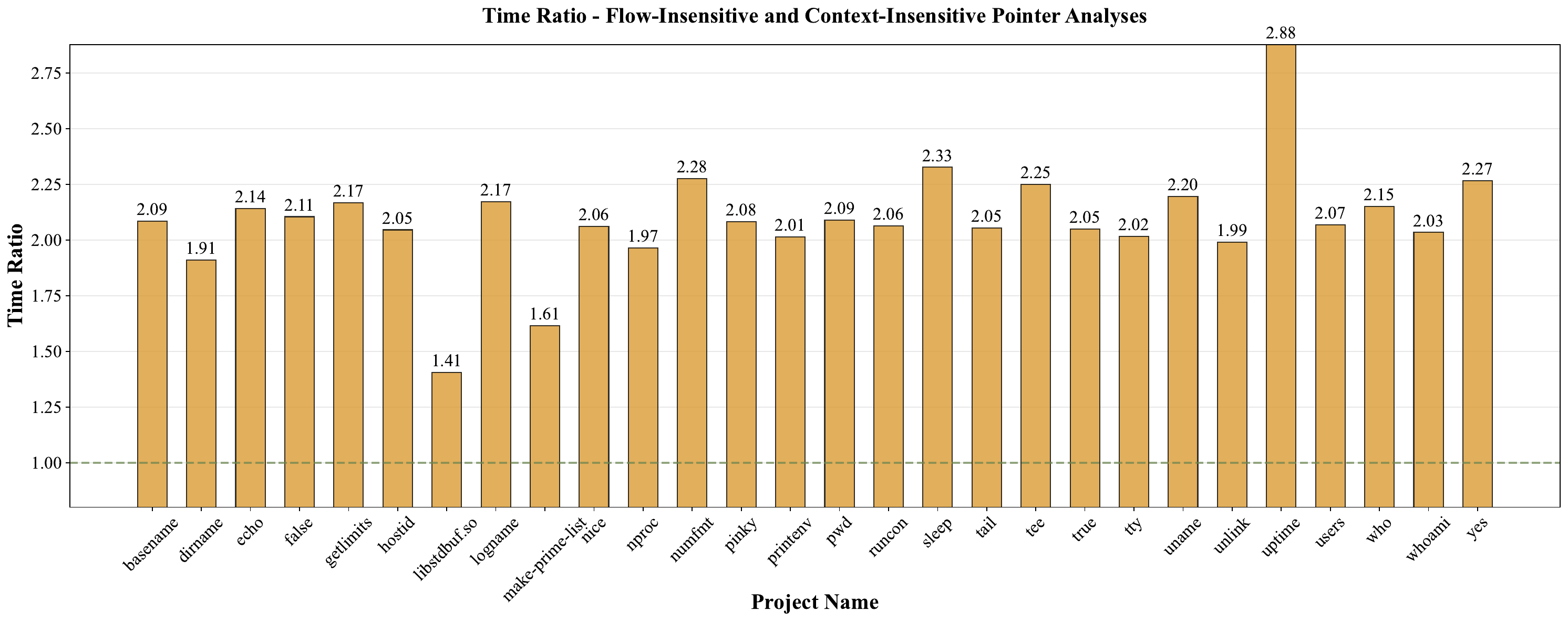}
  \caption{Runtime comparison of SVF-FICI vs.\ \ToolName-FICI. We report the runtime ratio $\text{Time}(\text{SVF})/\text{Time}(\text{\ToolName})$; values $>1$ indicate \ToolName\ is faster.}
  \label{fig:fici-runtime}
\end{figure}

\begin{figure}[t]
  \centering
   \includegraphics[width=\linewidth]{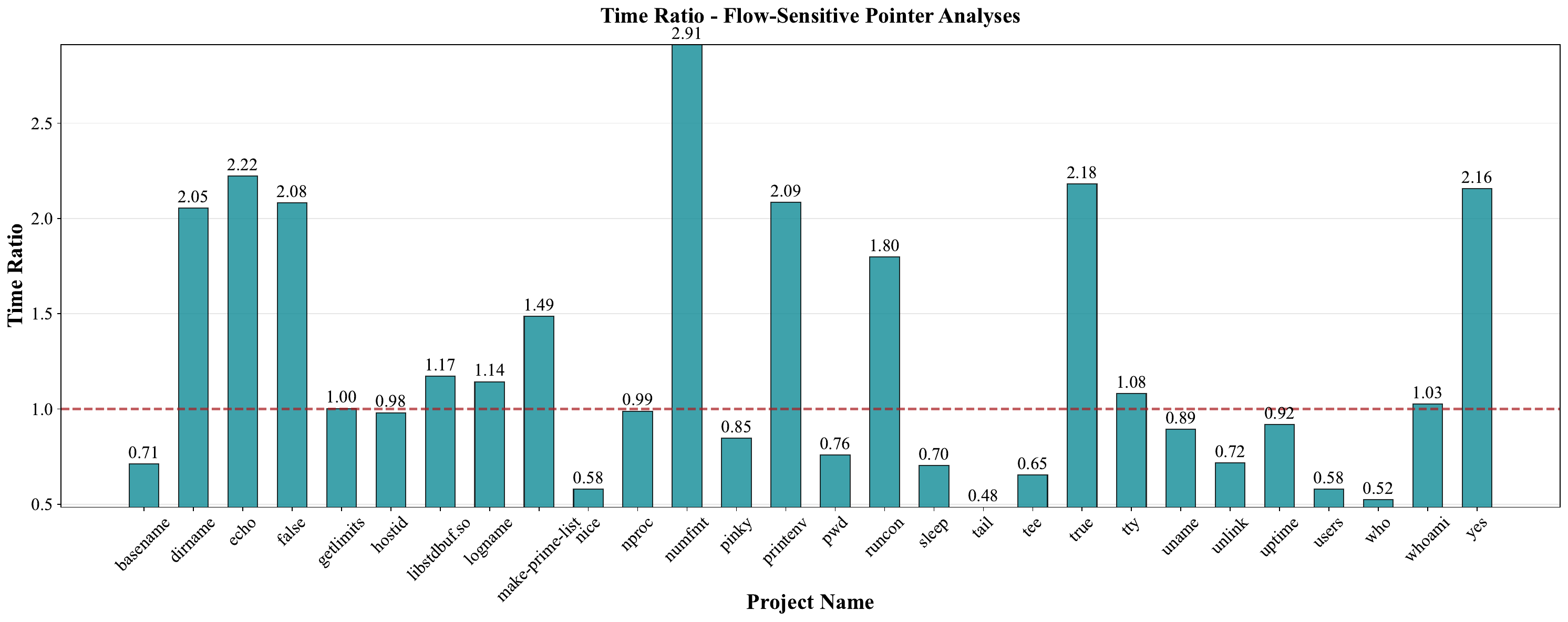}
  \caption{Runtime comparison of the flow-sensitive configurations: SVF-FSCI vs.\ \ToolName-FSCS. We report the runtime ratio $\text{Time}(\text{SVF})/\text{Time}(\text{\ToolName})$; values $>1$ indicate \ToolName\ is faster.}
  \label{fig:fscs-runtime}
\end{figure}

Figures~\ref{fig:fici-runtime} and~\ref{fig:fscs-runtime} summarize the runtime ratio of $\text{SVF}/\text{\ToolName}$. In the FICI setting (Figure~\ref{fig:fici-runtime}), \ToolName\ is faster on all 28 benchmarks, with ratios ranging from 1.41 to 2.88. In the FSCS setting (Figure~\ref{fig:fscs-runtime}), results are mixed: \ToolName-FSCS is faster on 13 benchmarks, slower on 14, and tied on one. The stronger configuration therefore has comparable aggregate performance but substantial variation across programs, including several significant slowdowns.

\smallskip
\noindent \textbf{Real-World Effectiveness}.
As discussed in \cref{subsec:uses}, \ToolName\ serves as the analysis substrate for several production-grade static analysis and fuzzing tools~\cite{HuangBeacon22,sun2024fast,HuangTitan24,OOPSLA22CFL}, which have helped detect memory-safety violations and logic errors in large-scale open-source software. When deployed in a continuous analysis pipeline, they have collectively discovered hundreds of previously unknown bugs, many of which were assigned CVEs. A curated subset of these findings is available at 
\url{https://tinyurl.com/3jwrev67}.
Notably, the directed fuzzing tool has been successfully deployed at Huawei and has detected hundreds more within the company.


\section{Conclusion}
\label{sec:conclusion}
We present \ToolName, a framework for pointer analysis that unifies a broad class of algorithms under a common interface. Its extensible architecture and clearly defined abstractions reduce the engineering effort required to implement new analyses.
\ToolName\ aims to accelerate both research prototyping and the development of practical tools for C/C++ programs.

We are actively pursuing several extensions to the framework:
\begin{itemize}
\item \emph{Path-Sensitive Pointer Analysis:} We are developing a variant of Andersen-style analysis that incorporates path sensitivity, building on the authors' recent work \cite{YaoFalcon24}.
\item \emph{More Built-in Analysis Clients:} We are expanding the suite of built-in downstream analyses, including support for the Weighted Pushdown System (WPDS)-based dataflow analysis framework~\cite{reps2005weighted} and the Newtonian program analysis framework~\cite{reps2016newtonian} for quantitative program analysis.
\item \emph{Dynamic Alias Analysis:} We are developing a dynamic alias analysis component to complement the static analyses and improve precision in hybrid static-dynamic analysis settings.
\end{itemize}

\section{Tool Availability}
\label{sec:tool-availability}

\ToolName\ is released as part of the Lotus program analysis framework~\cite{lotus2025}, which contains build instructions, usage guidance for the CLI tools, API documentation, and examples for downstream clients. 
An archived version of the artifact is
available at \url{https://doi.org/10.5281/zenodo.20928695}, and a demonstration video is
available at \url{https://www.youtube.com/watch?v=qabR7394TLM}.

\section*{Acknowledgments}
We sincerely thank the reviewers for their suggestions. This work is partially supported by the National Natural Science Foundation of China (62302434).

\bibliographystyle{ACM-Reference-Format}
\bibliography{bib/ref,bib/apps}

@inproceedings{kuderski2018teadsa,
  title={TeaDsa: Type-aware DSA-style pointer analysis for low level code},
  author={Kuderski, Jakub and L{\^e}, Nh{\^a}m and Gurfinkel, Arie and Navas, Jorge},
  booktitle={FMCAD},
  year={2018}
}

@article{ISSTA25Bit,
  author       = {Guangsheng Fan and
                  Liqian Chen and
                  Banghu Yin and
                  Wenyu Zhang and
                  Peisen Yao and
                  Ji Wang},
  title        = {Program Analysis Combining Generalized Bit-Level and Word-Level Abstractions},
  journal      = {Proc. {ACM} Softw. Eng.},
  volume       = {2},
  number       = {{ISSTA}},
  pages        = {663--685},
  year         = {2025},
  doi          = {10.1145/3728905}
}

@article{OOPSLA22CFL,
  author       = {Qingkai Shi and
                  Yongchao Wang and
                  Peisen Yao and
                  Charles Zhang},
  title        = {Indexing the extended Dyck-CFL reachability for context-sensitive
                  program analysis},
  journal      = {Proc. {ACM} Program. Lang.},
  volume       = {6},
  number       = {{OOPSLA2}},
  pages        = {1438--1468},
  year         = {2022},
  doi          = {10.1145/3563339},
}

@article{sun2024fast,
  title={Fast and Precise Static Null Exception Analysis With Synergistic Preprocessing},
  author={Sun, Yi and Wang, Chengpeng and Fan, Gang and Shi, Qingkai and Zhang, Xiangyu},
  journal={IEEE Transactions on Software Engineering},
  year={2024},
  publisher={IEEE}
}

@inproceedings{HuangTitan24,
	author = {Heqing Huang and Peisen Yao and Hung{-}Chun Chiu and Yiyuan Guo and Charles Zhang},
	title = {Titan : Efficient Multi-target Directed Greybox Fuzzing},
	booktitle = {{IEEE} Symposium on Security and Privacy, {SP} 2024, San Francisco, CA, USA, May 19-23, 2024},
	pages = {1849--1864},
	publisher = {{IEEE}},
	year = {2024},
	doi = {10.1109/SP54263.2024.00059}
}

@inproceedings{HuangBeacon22,
	author = {Heqing Huang and Yiyuan Guo and Qingkai Shi and Peisen Yao and Rongxin Wu and Charles Zhang},
	title = {{BEACON:} Directed Grey-Box Fuzzing with Provable Path Pruning},
	booktitle = {43rd {IEEE} Symposium on Security and Privacy, {SP} 2022, San Francisco, CA, USA, May 22-26, 2022},
	pages = {36--50},
	publisher = {{IEEE}},
	year = {2022},
	doi = {10.1109/SP46214.2022.9833751}
}

@inproceedings{gurfinkel2015seahorn,
  title={The SeaHorn verification framework},
  author={Gurfinkel, Arie and Kahsai, Temesghen and Komuravelli, Anvesh and Navas, Jorge A},
  booktitle={International Conference on Computer Aided Verification},
  pages={343--361},
  year={2015},
  organization={Springer}
}

@inproceedings{carter2016smack,
  title={SMACK software verification toolchain},
  author={Carter, Montgomery and He, Shaobo and Whitaker, Jonathan and Rakamari{\'c}, Zvonimir and Emmi, Michael},
  booktitle={Proceedings of the 38th International Conference on Software Engineering Companion},
  pages={589--592},
  year={2016}
}

@inproceedings{schubert2019phasar,
  title={Phasar: An inter-procedural static analysis framework for c/c++},
  author={Schubert, Philipp Dominik and Hermann, Ben and Bodden, Eric},
  booktitle={International Conference on Tools and Algorithms for the Construction and Analysis of Systems},
  pages={393--410},
  year={2019},
  organization={Springer}
}

@inproceedings{chalupa2020dg,
  title={DG: analysis and slicing of LLVM bitcode},
  author={Chalupa, Marek},
  booktitle={International Symposium on Automated Technology for Verification and Analysis},
  pages={557--563},
  year={2020},
  organization={Springer}
}

@inproceedings{liu2022pus,
  title={PUS: A fast and highly efficient solver for inclusion-based pointer analysis},
  author={Liu, Peiming and Li, Yanze and Swain, Brad and Huang, Jeff},
  booktitle={Proceedings of the 44th International Conference on Software Engineering},
  pages={1781--1792},
  year={2022}
}

@inproceedings{pereira2009wave,
  title={Wave propagation and deep propagation for pointer analysis},
  author={Pereira, Fernando Magno Quintao and Berlin, Daniel},
  booktitle={2009 International Symposium on Code Generation and Optimization},
  pages={126--135},
  year={2009},
  organization={IEEE}
}

@inproceedings{DOOP,
  title={Porting doop to souffl{\'e}: a tale of inter-engine portability for datalog-based analyses},
  author={Antoniadis, Tony and Triantafyllou, Konstantinos and Smaragdakis, Yannis},
  booktitle={Proceedings of the 6th ACM SIGPLAN International Workshop on State Of the Art in Program Analysis},
  pages={25--30},
  year={2017}
}

@inproceedings{SoopUp,
  title={Sootup: A redesign of the soot static analysis framework},
  author={Karakaya, Kadiray and Schott, Stefan and Klauke, Jonas and Bodden, Eric and Schmidt, Markus and Luo, Linghui and He, Dongjie},
  booktitle={International Conference on Tools and Algorithms for the Construction and Analysis of Systems},
  pages={229--247},
  year={2024},
  organization={Springer}
}

@incollection{Soot,
  title={Soot: A Java bytecode optimization framework},
  author={Vall{\'e}e-Rai, Raja and Co, Phong and Gagnon, Etienne and Hendren, Laurie and Lam, Patrick and Sundaresan, Vijay},
  booktitle={CASCON First Decade High Impact Papers},
  pages={214--224},
  year={2010}
}

@inproceedings{OPAL,
  title={Modular collaborative program analysis in OPAL},
  author={Helm, Dominik and K{\"u}bler, Florian and Reif, Michael and Eichberg, Michael and Mezini, Mira},
  booktitle={Proceedings of the 28th ACM Joint Meeting on European Software Engineering Conference and Symposium on the Foundations of Software Engineering},
  pages={184--196},
  year={2020}
}

@inproceedings{Taie,
  title={Tai-e: A developer-friendly static analysis framework for java by harnessing the good designs of classics},
  author={Tan, Tian and Li, Yue},
  booktitle={Proceedings of the 32nd ACM SIGSOFT International Symposium on Software Testing and Analysis},
  pages={1093--1105},
  year={2023}
}

@inproceedings{SeaDSA,
  title={Unification-based pointer analysis without oversharing},
  author={Kuderski, Jakub and Navas, Jorge A and Gurfinkel, Arie},
  booktitle={2019 Formal Methods in Computer Aided Design (FMCAD)},
  pages={37--45},
  year={2019},
  organization={IEEE}
}

@misc{lotus2025,
  title = {Lotus: A Versatile and Industrial-Scale Program Analysis Framework},
  author = {ZJU Programming Languages and Automated Reasoning Group},
  year = {2025},
  url = {https://github.com/ZJU-PL/lotus},
  note = {Program analysis framework built on LLVM}
}

@inproceedings{kroening2014cbmc,
  title={CBMC--C Bounded Model Checker: (Competition Contribution)},
  author={Kroening, Daniel and Tautschnig, Michael},
  booktitle={International Conference on Tools and Algorithms for the Construction and Analysis of Systems},
  pages={389--391},
  year={2014},
  organization={Springer}
}

@inproceedings{beyer2011cpachecker,
  title={CPAchecker: A tool for configurable software verification},
  author={Beyer, Dirk and Keremoglu, M Erkan},
  booktitle={International conference on computer aided verification},
  pages={184--190},
  year={2011},
  organization={Springer}
}

@article{YaoFalcon24,
	author = {Peisen Yao and Jinguo Zhou and Xiao Xiao and Qingkai Shi and Rongxin Wu and Charles Zhang},
	title = {Falcon: {A} Fused Approach to Path-Sensitive Sparse Data Dependence Analysis},
	journal = {Proc. {ACM} Program. Lang.},
	volume = {8},
	number = {{PLDI}},
	pages = {567--592},
	year = {2024},
	doi = {10.1145/3656400}
}

@article{reps2005weighted,
  title={Weighted pushdown systems and their application to interprocedural dataflow analysis},
  author={Reps, Thomas and Schwoon, Stefan and Jha, Somesh and Melski, David},
  journal={Science of Computer Programming},
  volume={58},
  number={1-2},
  pages={206--263},
  year={2005},
  publisher={Elsevier}
}

@inproceedings{reps2016newtonian,
  title={Newtonian program analysis via tensor product},
  author={Reps, Thomas and Turetsky, Emma and Prabhu, Prathmesh},
  booktitle={Proceedings of the 43rd Annual ACM SIGPLAN-SIGACT Symposium on Principles of Programming Languages},
  pages={663--677},
  year={2016}
}

@inproceedings{tan2026simd,
	author = {Zhaoyang Tan and Peisen Yao and Kui Ren},
	title = {{SIMD}-Accelerated Sparse Bit-Vectors for Pointer Analysis},
	booktitle = {Proceedings of the 41st {IEEE}/{ACM} International Conference on Automated Software Engineering, {ASE} 2026},
	year = {2026}
}

@inproceedings{wang2017partitioned,
	author       = {Wei Wang and Clark W. Barrett and Thomas Wies},
	editor       = {Ahmed Bouajjani and David Monniaux},
	title        = {Partitioned Memory Models for Program Analysis},
	booktitle    = {Verification, Model Checking, and Abstract Interpretation - 18th International Conference, {VMCAI} 2017, Paris, France, January 15-17, 2017, Proceedings},
	series       = {Lecture Notes in Computer Science},
	volume       = 10145,
	year         = 2017,
	bibsource    = {dblp computer science bibliography, https://dblp.org}
}

@inproceedings{gurfinkel2017context,
	title        = {A context-sensitive memory model for verification of C/C++ programs},
	author       = {Gurfinkel, Arie and Navas, Jorge A},
	booktitle    = {International Static Analysis Symposium},
	series       = {SAS '17},
	year         = 2017,
	organization = {Springer}
}

@inproceedings{li2016program,
	title        = {Program tailoring: Slicing by sequential criteria},
	author       = {Li, Yue and Tan, Tian and Zhang, Yifei and Xue, Jingling},
	booktitle    = {30th European Conference on Object-Oriented Programming (ECOOP 2016)},
	series       = {Leibniz International Proceedings in Informatics (LIPIcs)},
	issn         = {1868-8969},
	year         = 2016,
	volume       = 56,
	editor       = {Shriram Krishnamurthi and Benjamin S. Lerner},
	address      = {Dagstuhl, Germany},
	urn          = {urn:nbn:de:0030-drops-61092}
}

@article{reps1998program2,
	title        = {Program analysis via graph reachability},
	author       = {Reps, Thomas},
	journal      = {Information and software technology},
	volume       = 40,
	number       = 11,
	year         = 1998
}

@article{smaragdakis2015pointer,
	title        = {Pointer analysis},
	author       = {Smaragdakis, Yannis and Balatsouras, George and others},
	journal      = {Found. Trends Program. Lang.},
	issue_date   = {4 2015},
	volume       = 2,
	number       = 1,
	month        = apr,
	year         = 2015,
	issn         = {2325-1107},
	address      = {Hanover, MA, USA}
}

@inproceedings{hind2001pointer,
	title        = {Pointer analysis: Haven't we solved this problem yet?},
	author       = {Hind, Michael},
	booktitle    = {Proceedings of the 2001 ACM SIGPLAN-SIGSOFT Workshop on Program Analysis for Software Tools and Engineering},
	series       = {PASTE '01},
	year         = 2001,
	address      = {New York, NY, USA}
}

@inproceedings{hardekopf2007ant,
	title        = {The ant and the grasshopper: fast and accurate pointer analysis for millions of lines of code},
	author       = {Hardekopf, Ben and Lin, Calvin},
	booktitle    = {Proceedings of the 28th ACM SIGPLAN Conference on Programming Language Design and Implementation},
	series       = {PLDI '07},
	year         = 2007,
	address      = {New York, NY, USA}
}

@inproceedings{sui2016svf,
	title        = {SVF: interprocedural static value-flow analysis in LLVM},
	author       = {Sui, Yulei and Xue, Jingling},
	booktitle    = {Proceedings of the 25th International Conference on Compiler Construction},
	series       = {CC 2016},
	year         = 2016,
	address      = {New York, NY, USA}
}

@article{pearce2007efficient,
	title        = {Efficient field-sensitive pointer analysis of C},
	author       = {Pearce, David J and Kelly, Paul HJ and Hankin, Chris},
	journal      = {ACM Trans. Program. Lang. Syst.},
	issue_date   = {November 2007},
	volume       = 30,
	number       = 1,
	month        = nov,
	year         = 2007,
	issn         = {0164-0925},
	articleno    = 4,
	address      = {New York, NY, USA}
}

@inproceedings{hardekopf2007exploiting,
	title        = {Exploiting pointer and location equivalence to optimize pointer analysis},
	author       = {Hardekopf, Ben and Lin, Calvin},
	booktitle    = {Proceedings of the 14th International Conference on Static Analysis},
	series       = {SAS'07},
	year         = 2007,
	address      = {Berlin, Heidelberg}
}

@inproceedings{lattner2007making,
	title        = {Making context-sensitive points-to analysis with heap cloning practical for the real world},
	author       = {Lattner, Chris and Lenharth, Andrew and Adve, Vikram},
	booktitle    = {Proceedings of the 28th ACM SIGPLAN Conference on Programming Language Design and Implementation},
	series       = {PLDI '07},
	year         = 2007,
	address      = {New York, NY, USA}
}

@inproceedings{fink:typestate:issta,
	author       = {Fink, Stephen and Yahav, Eran and Dor, Nurit and Ramalingam, G. and Geay, Emmanuel},
	title        = {Effective Typestate Verification in the Presence of Aliasing},
	booktitle    = {Proceedings of the 2006 International Symposium on Software Testing and Analysis},
	series       = {ISSTA '06},
	year         = 2006,
	address      = {New York, NY, USA}
}

@inproceedings{Orso:2004:ECD:998675.999453,
	author       = {Orso, Alessandro and Apiwattanapong, Taweesup and Law, James and Rothermel, Gregg and Harrold, Mary Jean},
	title        = {An Empirical Comparison of Dynamic Impact Analysis Algorithms},
	booktitle    = {Proceedings of the 26th International Conference on Software Engineering},
	series       = {ICSE '04},
	year         = 2004,
	address      = {Washington, DC, USA}
}

@inproceedings{sui2012static,
	author       = {Yulei Sui and Ding Ye and Jingling Xue},
	editor       = {Mats Per Erik Heimdahl and Zhendong Su},
	title        = {Static memory leak detection using full-sparse value-flow analysis},
	booktitle    = {International Symposium on Software Testing and Analysis, {ISSTA} 2012, Minneapolis, MN, USA, July 15-20, 2012},
	year         = 2012,
	bibsource    = {dblp computer science bibliography, https://dblp.org}
}

@inproceedings{shi2018pinpoint,
	title        = {Pinpoint: fast and precise sparse value flow analysis for million lines of code},
	author       = {Shi, Qingkai and Xiao, Xiao and Wu, Rongxin and Zhou, Jinguo and Fan, Gang and Zhang, Charles},
	booktitle    = {Proceedings of the 39th ACM SIGPLAN Conference on Programming Language Design and Implementation},
	series       = {PLDI 2018},
	year         = 2018,
	address      = {New York, NY, USA}
}

@inproceedings{sridharan2007thin,
	title        = {Thin slicing},
	author       = {Sridharan, Manu and Fink, Stephen J and Bodik, Rastislav},
	booktitle    = {Proceedings of the 28th ACM SIGPLAN Conference on Programming Language Design and Implementation},
	series       = {PLDI '07},
	year         = 2007,
	address      = {New York, NY, USA}
}

@inproceedings{tripp2009taj,
	title        = {TAJ: effective taint analysis of web applications},
	author       = {Tripp, Omer and Pistoia, Marco and Fink, Stephen J and Sridharan, Manu and Weisman, Omri},
	booktitle    = {Proceedings of the 30th ACM SIGPLAN Conference on Programming Language Design and Implementation},
	series       = {PLDI '09},
	year         = 2009,
	address      = {New York, NY, USA}
}

@inproceedings{hind2000pointer,
	title        = {Which pointer analysis should I use?},
	author       = {Hind, Michael and Pioli, Anthony},
	booktitle    = {Proceedings of the 2000 ACM SIGSOFT International Symposium on Software Testing and Analysis},
	series       = {ISSTA '00},
	year         = 2000,
	address      = {New York, NY, USA}
}

@inproceedings{lattner2004llvm,
	title        = {LLVM: A compilation framework for lifelong program analysis \& transformation},
	author       = {Lattner, Chris and Adve, Vikram},
	booktitle    = {Proceedings of the International Symposium on Code Generation and Optimization: Feedback-directed and Runtime Optimization},
	series       = {CGO '04},
	year         = 2004,
	address      = {Washington, DC, USA}
}

@inproceedings{WALA,
	title        = {T.J. Watson libraries for analysis},
	author       = {WALA, http://wala.sf.net/}
}

\end{document}